\documentclass[multphys,vecphys]{svmult}

\usepackage{makeidx}         
\usepackage{graphicx}        
\usepackage{multicol}        
\makeindex             

\begin{document}

\title*{A Century of Cosmology}
\author{Edward L. Wright\inst{1}}
\institute{UCLA Astronomy, PO Box 951547, Los Angeles, CA 90095-1547, USA
\texttt{wright@astro.ucla.edu}}
%
%
\maketitle

\begin{abstract}
In the century since Einstein's anno mirabilis of 1905, our concept of
the Universe has expanded from Kapteyn's flattened disk of stars only
10 kpc across to an observed horizon about 30 Gpc across that is only a
tiny fraction of an immensely large inflated bubble.  The expansion of
our knowledge about the Universe, both in the types of data and the
sheer quantity of data, has been just as dramatic.  This talk will
summarize this century of progress and our current understanding of the
cosmos.

\end{abstract}

\keywords{cosmic microwave background, cosmology: observations, 
early universe, dark matter}

\section{Introduction}\label{intro}

When the COBE DMR results were announced in 1992, Hawking was quoted in
The Times stating that "It was the discovery of the century, if not of
all time."  But the progress in cosmology in the last century has been
tremendous, going far beyond the anisotropy of the cosmic microwave
background.  A century ago the ``Structure of the Universe'' meant the
patterns of stellar number counts and proper motions that delineated
the discus-shaped distribution of observed stars [1].  The true scale
of the Milky Way and the nature of the extragalactic nebulae were yet
to be determined.  As late as 1963 people could say that there were
only 2.5 facts in cosmology [2]: 1) the sky is dark at night, 2) the
redshifts of galaxies show a pattern consistent with a general
expansion of the Universe, and 2.5) the Universe has evolved over
time.  In 1963 the controversy between the Steady State [3,4] and the
Big Bang [5,6] models of cosmology was still quite active, so the last
item in the list was only a half-fact.

\section{Einstein and $\lambda$}\label{early}

Once Einstein developed general relativity, giving a theory for
classical gravity, he worked out a cosmological model [7] using what
was then known about the Universe.  Einstein assumed that the Universe
had to be homogeneous, since even if the matter were confined to a
finite region initially, the action of gravitational scattering would
lead to stars being ejected from the initial distribution.  Since the
solution of Poisson's equation for a uniform density extending to
infinity is not well defined, Einstein considered modifying Newtonian
gravity by adding a $\lambda$ term, giving $\nabla^2\phi -\lambda\phi =
4\pi G\rho$, which has the constant solution $\phi = -4\pi
G\rho/\lambda$ for constant density.  This modified Newtonian gravity
has a short range compared to the infinite range inverse square law
behavior of normal gravity.  But in General Relativity the $\lambda$
term had to be multiplied not by $\phi$, which is not covariant, but
rather by the metric $g_{\mu\nu}$ which contains $\phi$ as $g_{00}
\approx 1+2\phi/c^2$ in the Newtonian approximation.  Einstein found
that for a uniform distribution of matter the geometry of space was
that of the 3-sphere $S^3$ (the surface of a 4 dimensional ball), and
that the addition of the $\lambda$ term could compensate for the
tendency for the Universe to collapse.

This static, spherical, homogeneous and isotropic Einstein Universe was
not compatible with a solution to Olbers' Paradox.  The stars in the
Universe were emitting light, and this light would circulate around the
spherical Universe and never be lost.  As the stars continued to emit
light, the Universe would become brighter and brighter.  In addition to
not solving Olbers' Paradox, Einstein's static Universe was only a
unstable equilibrium point between a collapsing model and an infinitely
expanding model.  After the redshift of distant galaxies was discovered
[8], Einstein referred to the introduction of $\lambda$ as his greatest
blunder.

Other cosmological models were developed as well.  The de Sitter
Universe used only $\lambda$ and had no matter [9].  It has a redshift
growing with distance, consistent with the Hubble Law, and this metric
is now recognized as a homogeneous and isotropic Euclidean space
(``flat'' space means a 3 dimensional Euclidean geometry) that is
expanding exponentially with time.  The metric of the Steady State
model is exactly the same as the de Sitter metric, but since the Steady
State model has both matter and the continuous creation of new matter,
the $\lambda$ term was replaced with a $C$-field that made the matter
plus $C$-field in the Steady State be equivalent to the pure vacuum
energy of the de Sitter space..

Friedmann introduced models with matter that expanded from an initial
singularity [10].  These models show a redshift proportional to
distance which is consistent with the Hubble Law.

\section{Big Bang vs. Steady State}\label{BBvsSS}

After World War II Gamow tried to use the new knowledge about nuclear
physics in a cosmological context.  He and his students considered
first a Universe full of neutrons that expanded and decayed. But they
changed to a Universe initially filled with a hot dense medium about
equally split between neutrons and protons.  As the Universe expanded
and cooled, heavier elements would be formed by the successive addition
of neutrons.  In this model, nuclei with high neutron capture
cross-sections would be rapidly converted into heavier nuclei, and
would thus be rare in the current Universe.  Indeed, a roughly inverse
relation between abundance and neutron capture cross section is
observed.  The time vs temperature during the cooling is related to the
matter to radiation ratio in the Universe, and then by estimating the
current density of matter, it was possible to estimate the current
temperature of Universe [6] as 1 K or 5 K.

In this model, the current Universe is more or less curvature dominated
so the ratio $\Omega = \rho/\rho_{crit}$ is $<< 1$ and therefore the
age of the Universe is $t_\circ \approx 1/H_\circ =
978\;\mbox{Gyr}/H_\circ\;\mbox{in km/sec/Mpc}$.  Since the value of the
Hubble constant given by Hubble was $\approx 500\;\mbox{km/sec/Mpc}$
the age of the Universe was about 2 Gyr, which was too short according
to the radioactive dating of the Earth.  In the Steady State model the
scale factor of the Universe is an exponential function of time, $a(t)
= \exp(H(t-t_\circ))$, and thus the Hubble constant is actually a
constant and the age of the Universe is infinite.  But the average age
of the matter in the Universe is in fact quite short: $\langle \tau
\rangle = 1/3H \approx 700\;\mbox{Myr}$ for Hubble's value of the
Hubble constant.  Taking 6 Gyr as a minimum age for the Milky Way based
on the radioactive dating of the Earth and adding time needed to form
the galaxy and stars, the probability that a random piece of the
Universe would be that old or older was only $e^{-9} \approx 10^{-4}$
in the Steady State model but this was still better than the zero
probability in the Big Bang model.

\section{Discovery and Non-discovery of the CMB}\label{CMB}

The first evidence [11] for the CMB was a rather inconspicuous
interstellar absorption line in the spectrum of the hot, rapidly
rotating star $\zeta$ Oph.  This line was identified with the R(1) line
of the cyanogen radical, CN.  It was rather unusual, since it arises
from a rotationally excited state of CN.  Given the low density of the
interstellar medium, ions and molecules in the ISM spend almost all of
the time in the ground state.  The excitation temperature [12] of the
rotational transition based on this first CN data was 2.3 K,  but its
cosmological significance was widely ignored.  Nobel Prize winner
Herzberg stated: ``From the intensity ratio of the lines with K=0 and
K=1 a rotational temperature of $2.3^\circ$ K follows, which has of
course only a very restricted meaning.'' But it is not true that the
cosmological significance was completely ignored.  In fact Hoyle, in a
review [13] of a book by Gamow \& Critchfield, wrote that ``the authors
use a cosmological model in direct conflict with more widely accepted
results.  The age of the Universe is this model is appreciably less
than the agreed age of the Galaxy.  Moreover it would lead to a
temperature of the radiation at present maintained throughout the whole
of space much greater than McKellar's determination for some regions
within the Galaxy.'' In making this statement Hoyle ignored the careful
and explicit calculations of $T_\circ$ contained in a refereed article
[6], which were perfectly compatible with the CN temperature measured
by McKellar.  I find it remarkable that none of the parties involved
thought to follow up this possibility of a decisive test of the Big
Bang vs. Steady State.

As a result, the actual discovery of the CMB was left to Penzias \&
Wilson, who were quite dedicated to finding the source of the excess
noise they saw in their low-noise microwave receiver.  Within 7 months
of the announcement of Penzias \& Wilson's result,
the brightness temperature of the CMB at the 2.6 mm wavelength of the
CN rotational transition had been shown to be the same as that measured
at 7.4 cm by Penzias \& Wilson.  And thus Gamow, Alpher, Herman and
Hoyle all missed the Nobel Proze.

Bob Dicke also narrowly missed the Nobel Prize.  He invented the Dicke
radiometer used in all direct measurements of the CMB spectrum, and
during World War II came within a factor of ten of discovering the CMB,
even while working from a sea level location in Florida [14].  While
building a radiometer with a cold load to specifically search for the
CMB, Dicke, Roll, Peebles \& Wilkinson heard from Penzias \& Wilson
about their work.

\section{Nucleosynthesis}\label{BBNS}

Since Gamow's motivation for the Big Bang model was the origin of the
chemical elements, it is instructive to see how the Big Bang and Steady
State models fare on isotopic abundances.  One cannot make isotopes
heavier than ${}^4He$ by the sequential addition of neutrons in the Big
Bang because there is no nucleus with atomic weight $A = 5$ that is
even slightly stable.  Thus the Big Bang, which set out to explain the
abundances of the elements from hydrogen to uranium ended up only able
to produce the elements from hydrogen to helium, with a sprinkling of
lithium.

The Steady State model, on the other hand, proposed that matter was
continuously created in the form of hydrogen, and that all heavier
elements were created in stars.  The triple-$\alpha$ reaction, ${}^4He
+ {}^4He + {}^4He \rightarrow {}^{12}C$, can run in stars because
conditions of high density and high temperature persist for a long
time.  Thus all the elements can be produced in stars, starting by
fusing hydrogen into helium.  Stars produce about 1 gram of
elements heavier than helium (``metals'' to an astronomer) for each 3
grams of helium.  But the average helium to metals ratio is about 12 to
1, and in low metallicity stars the ratio is even higher.  Thus the
Steady State model fails to produce enough helium, leading to the
``helium problem''.  A proposed solution [15] to this problem was to
have the ongoing creation of matter in the Steady State model occur
sporadically in a number of ``little bangs'' that produce a mixture of
hydrogen and helium.

The current model uses a combination of Big Bang nucleosynthesis, which
produces most of the helium, and stellar nucleosynthesis, which
produces the metals and some helium.  Nuclear reactions during the Big
Bang, starting from a mixture of protons and neutrons in thermal
equilibrium at $t < 1$\,sec after the Big Bang and $T > 10^{10}$\,K,
produce the deuterium ($D$), ${}^3He$, ${}^4He$ and ${}^7Li$ seen in
material that has not been processed through stars.  Stars can destroy
$D$ and ${}^7Li$, and generate more ${}^4He$.  The predicted abundances
depend on the number of neutrino species and the baryon to photon
ratio.  The number of neutrino species primarily controls the ${}^4He$
abundance, and appears to be 3, consistent with determinations based on
the decay width of the Z boson.  The baryon to photon ratio controls
the D:H ratio and the  ${}^7Li$ abundance.  The baryon to photon ratio
consistent with the D:H ratio seen in high redshift quasar absorption
line systems appears to predict a higher ${}^7Li$ abundance than that
observed in a certain class of stars that has been thought not to have
destroyed lithium in their surface layers.  But since stars certainly
do destroy lithium in their interiors this discrepancy is not too
serious.

\section{CMB Anisotropy}\label{DT}

The CMB was found to be remarkably isotropic.  This provided strong
evidence that the simple Friedmann-Robertson-Walker metrics, adopted as
a useful approximation, were actually quite good representations of the
real Universe.  While galaxy counts in different directions as a
function of brightness had already demonstrated that the Universe was
homogeneous and isotropic on large scales, it was still possible in
1967 to propose 10\% inhomogeneities leading to 1\% anisotropies in the
CMB [16].  The first detection of the CMB anisotropy was at the 0.1\%
level [17,18], and it was soon in textbooks [19] as due to the motion
of the Solar System relative to the Universe.  The alleged
``discovery'' of the dipole anisotropy by the U2 experiment [20] was
published 6 years after the textbook.  Anisotropy other than this
dipole term was not detected until 1992, by the COBE DMR [21,22] at the
0.001\% level.

The low level of the anisotropy seen by COBE was strong evidence for
the existence of dark matter.  Dark matter can start to collapse as
soon as the matter density exceeds the radiation density, while
baryonic matter is frozen to the photons until recombination.   Thus
there is more growth for structures in dark matter dominated models,
and thus the currently observed large scale structure can be generated
starting from smaller seeds and hence smaller CMB anisotropies [23].
But the ratio of fluctuations at supercluster scales to the
fluctuations at cluster scales required a modest reduction in the small
scale power that could be supplied by either an open Universe model
(OCDM), a model with a mixture of hot and cold running dark matter
(mixed dark matter, or MDM), or by a model dominated by a cosmological
constant ($\Lambda$CDM) [22].

Detailed calculations of cold dark models (CDM) showed acoustic oscillations
in the amplitude of the anisotropy as function of angular scale [24].  These 
oscillations were caused by an interference between the fluctuations in the
dark matter, which has zero pressure and thus zero sound speed, and the 
baryon-photon fluid which has a sound speed near 170,000 km/sec.  The
angular scale of the main peak of the angular power spectrum depends on two
parameters: whether the Universe is open, flat or closed ($\Omega_{tot}$),
and the amount of vacuum energy density ($\Omega_\Lambda$).

The first observational evidence for the main acoustic peak came from a
collection of data from different experiments [25].  By 2003, the {\sl
WMAP} experiment [26] had measured the position of the first peak to an
accuracy of better than 0.5\% [27].  The result requires the Universe
to lie along a line segment in the $\Omega_{tot}$ vs $\Omega_\Lambda$
plane, with allowable models lying between a flat $\Lambda$CDM model
having $\Omega_{tot} = 1$ and $\Omega_\Lambda = 0.73$ and a closed
``super-Sandage'' model with $\Omega_{tot} = 1.3$  and  $\Omega_\Lambda
= 0$.  This model is referred to as super-Sandage because it has a
Hubble constant of $H_\circ = 32$\,km/sec/Mpc.  The ratio of the first
peak amplitude to second peak amplitude and to the valley between the
peaks determined the ratio of the dark matter to baryon density and the
baryon to photon density.  The photon density is well measured by FIRAS
on COBE, so the physical matter density $\Omega_m h^2 = 0.135 \pm 10\%$
is determined.  Thus a value of $\Omega_m = \Omega_{tot} -
\Omega_\Lambda$ also determines a Hubble constant, and the
super-Sandage model has $H_\circ = 32$\,km/sec/Mpc.

Thus the CMB anisotropy data alone cannot tell whether the Universe is flat
or not, and cannot say that the cosmological constant is non-zero. This
comes from the fact that the CMB anisotropy power spectrum is generated
at recombination, when $z = 1089$, and at high redshifts the cosmological
constant is a negligible contribution to the overall density.  Other
data are needed to verify the existence of the cosmological constant.

\section{Supernovae}\label{SN}

This other data was provided by observations of Type Ia supernovae.
A definite correlation between the decay rate and peak luminosity of Ia SNe
was seen [28],  and using this calibration it was possible to pin down the
acceleration of the Universe.  This acceleration is usually denoted by
the deceleration parameter, $q_\circ = \Omega_m/2 - \Omega_\Lambda$.
If the expansion of the Universe is accelerating, it was slower in the past,
and thus a larger time is needed to reach a given expansion ratio or
redshift.  With the larger travel time comes a larger distance, so 
distant supernovae appear fainter in accelerating models.  Based on the
SNe data, the Universe is definitely accelerating, so $q_\circ$ is
negative.  But the supernova data could be affected by systematic errors.
In particular, evolution of the zero-point of the supernova decay rate
{\it vs.} peak luminosity calibration can in principle match any cosmological
model.  In fact, a very simple exponential in cosmic time evolution
in an Einstein - de Sitter Universe matches the supernova data very well, 
and is actually a slightly better fit to the data than the flat 
$\Lambda$CDM model with no evolution [31].  Since there is no way to
rule out evolution with supernova data alone, the existence of the
cosmological constant needs to be confirmed using other data or a
combination of other data.

There are many other datasets that do confirm the acceleration first
seen in the supernova data.  The CMB anisotropy in combination with the
Hubble constant data require an accelerating, close-to-flat Universe,
as does the combination of CMB data and the peak of the large scale
structure power spectrum $P(k)$, or the correlation of the CMB
temperature fluctuations with superclusters via the late-integrated
Sachs-Wolfe effect.

\section{Search for Two Numbers}\label{twonum}

In the February 1970 {\it Physics Today}, Sandage published an article
[32] titled ``Cosmology: A search for Two Numbers''.  At that time,
since the cosmological constant had fallen out of favor, the two
numbers being sought were the Hubble constant $H_\circ = \dot{a}/a$ and
the deceleration parameter $q_\circ = -\ddot{a}a/\dot{a}^2$, where
$a(t)$ is the scale factor.  It is historically interesting that
Sandage gave $H_\circ = 80\; \mbox{km/sec/Mpc}\pm$ a factor of 1.6 for
the Hubble constant, in view of the later distance scale controversy.
But his value for the deceleration parameter, $q_\circ = 1.2 \pm 0.4$,
is far from the currently accepted $q_\circ = -0.6$.

\begin{figure}
\centering
\includegraphics[height=10cm]{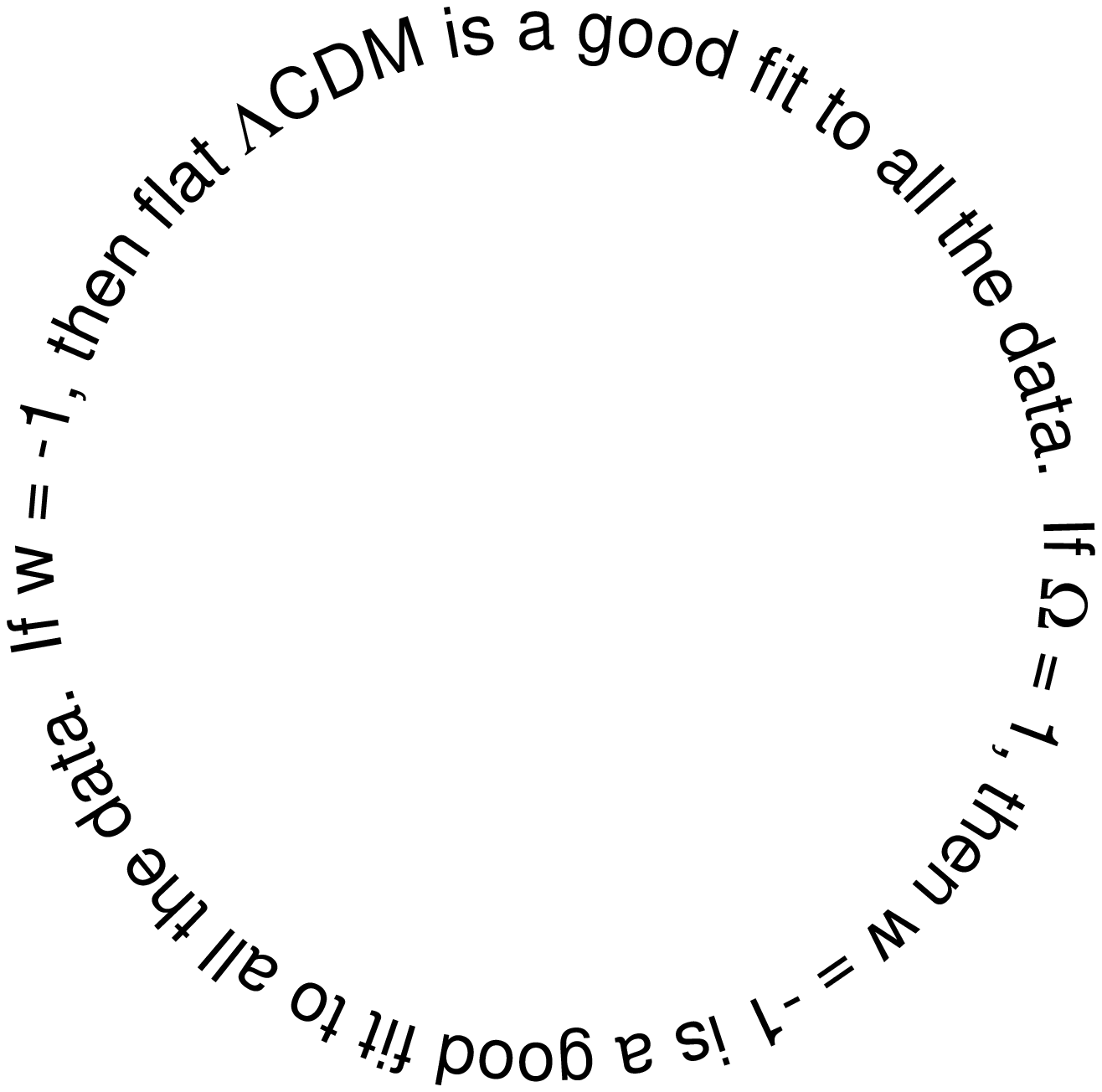}
%
%
\caption{The circular argument popular among current searches for
$w$ and $w^\prime$.  Models fits should always allow $\Omega_{tot}$ to
be a free parameter.\label{fig:circular}}
\end{figure}

\begin{figure}
\centering
\includegraphics[height=4cm]{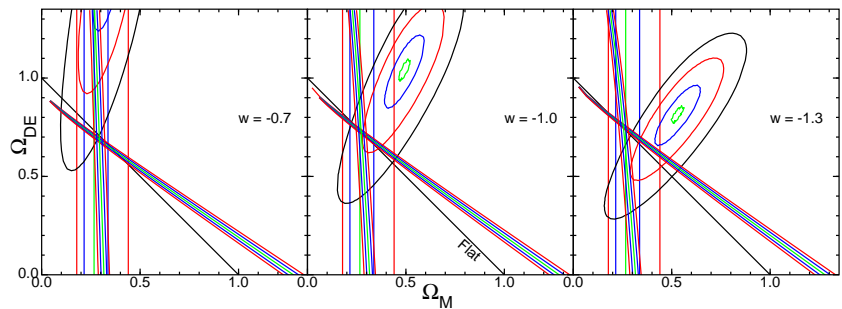}
\caption{Constraints on the dark energy density and the matter density
from four relatively precise cosmological measurements for three
different values of the dark energy equation of state.
The ellipses show $\Delta \chi^2 = 
0.1,\;1,\;4\;\&\;9$ for my fits to the supernova data [29],
while the lines show 
$-2\sigma,\;-1\sigma,\;0,\;1\sigma,\; \&\;2\sigma$ values for $H_\circ$
[33] (vertical), 
baryonic oscillations in the SDSS [34] (not quite vertical),
and the CMB acoustic peak angular scale [27] (inclined). 
The concordance at $w = -1$ is gradually lost for other values of
$w$.\label{fig:w}}
\end{figure}

The current cosmological literature is again seeking two numbers, but a
different set of two numbers.  These are the equation of state
parameter $w = P/\rho c^2$ and its time derivative $w^\prime$.  $w$ is
exactly $-1$ for a cosmological constant, but will be different for
models involving evolving scalar fields.  If $w$ is not exactly  $-1$
then it will interact with matter fluctuations via gravity, and thus
the dark energy will be a function of both space and time or redshift.
But since the Universe is almost homogeneous and isotropic, the average
of $w$ and $w^\prime$ over space should be a good description of the
evolution of dark energy.

However, there is a very strong tendency among theorists to assume the
Universe is flat when seeking $w$ and $w^\prime$.  This is a logical
error, since the evidence for a flat Universe comes from the agreement
of the concordance $\Lambda$CDM model with all the data.  But the
concordance $\Lambda$CDM model has $w = -1$ and $w^\prime = 0$
exactly.  If $w$ and $w^\prime$ are allowed to vary, then the evidence
for a flat Universe must be re-evaluated.  Limits on $w$ and $w^\prime$
are only valid when a simultaneous fit for all relevant parameters is
done.  And when fitting to the CMB data, the spatial variations in the
dark energy density should be included even though they are
\cal{O}$(10^{-5})$, since the CMB $\Delta T/T$ is of the same order.

\section{Discussion}\label{sec:disc}

The progress in cosmology over the past century has been astronomical.
We have gone from one fact in 1905 to hundreds of observed facts in
2005.

In terms of the mass of the known Universe the progress is even
greater.  In 1905 Kapteyn might have given the mass of the Universe as
$10^9\; M_\odot$.  Today the mass of the Universe is much larger than
the mass of the Hubble volume $M = (4\pi/3) \rho_{crit} (c/H_\circ)^3
= 0.5 c^3 /(G H_\circ) = 4.4 \times 10^{22}\;M_\odot$ so we can claim
to have discovered more than 44 trillion times more of the Universe
than was known in 1905.  But we have also found that 95\% of the
density of the Universe is mysterious dark matter or dark energy.

Cosmologists today are working on problems that could hardly have been
defined in 1905, but they are fortunate in having a large and growing
body of precise observations with which to test their speculative
constructs.  Further observations of the CMB, large scale supernovae
surveys, weak lensing and baryon oscillations will all provide major
new datasets in the next century, and future progress in cosmology is
assured.

\printindex


\begin{thebibliography}{99.}

\bibitem{1}  J.~Kapteyn, 1914, JRASC, 8, 145.

\bibitem{2}  M.~Longair, 1993, QJRAS, 34, 157.

\bibitem{3}  F.~Hoyle, 1948, MNRAS, 108, 372.

\bibitem{4}  H.~Bondi and T.~Gold, 1948, MNRAS, 108, 252.

\bibitem{5} G.~Gamow, 1948, Physical Review, 74, 505.

\bibitem{6}  R.~Alpher and R.~Hermann, 1949, Physical Review, 75, 1089.

\bibitem{7}  A.~Einstein, 1917, Sitzung Derichte per Preussischen Akad.
d. Wiss., 1917, 142,

\bibitem{8}  E.~Hubble, 1929, PNAS, 15, 168.

\bibitem{9}  W.~de Sitter, 1917, MNRAS, 78, 3.

\bibitem{10}  A.~Friedmann, 1922, Zeitschrift f\"ur Physik, 21, 326.

\bibitem{11}  W. Adams, 1941, ApJ, 93, 11.

\bibitem{12}  G. Herzberg, 1950, ``Spectra of Diatomic Molecules'',
(New York: Van Nostrand Reinhold)

\bibitem{13}  F. Hoyle, 1950, The Observatory, 70, 194-197.

\bibitem{14}  R. Dicke, R. Behringer, R. Kyhl \& A. Vane, 1946, 
Physical Review, 70, 340-348.

\bibitem{15}  F. Hoyle \& R. Tayler, 1964, Nature, 203 1108.

\bibitem{16}  R. Sachs and A. Wolfe, 1967, ApJ, 147, 73.

\bibitem{17}  E. Conklin, 1969, Nature, 222, 971-972.

\bibitem{18}  P. Henry, 1971, 231, 516.

\bibitem{19}  P. Peebles, 1971, ``Physical Cosmology'', (Princeton: Princeton
University Press)

\bibitem{20}  G. Smoot, Gorenstein, M. \& R. Muller, 1977, PRL, 39, 898.

\bibitem{21}  G. Smoot et al., 1992, ApJL, 396, L1.

\bibitem{22}  E. Wright et al., 1992, ApJL, 396, L13.

\bibitem{23}  P. Peebles, 1982, ApJ, 263, L1-L5.

\bibitem{24}  J. Bond \& G. Efstathiou, 1987, MNRAS, 226, 655-687.

\bibitem{25}  D. Scott, J. Silk, \& M. White, 1995, Science, 268, 829-835.

\bibitem{26}  C. Bennett et al., 1993, ApJS, 148, 1.

\bibitem{27}  L. Page et al., 2003, ApJS, 148, 233.

\bibitem{28}  M. Phillips, 1993, ApJL, 413, L105-L108.

\bibitem{29}  A. Riess et al., 2004, ApJ,  607, 665-687.

\bibitem{30}  S. Perlmutter et al., 1999, ApJ, 517, 565-586.

\bibitem{31}  E. Wright, 2002, astro-ph/0201196.

\bibitem{32}  A. Sandage, 1970, Physics Today, 23, 34.

\bibitem{33}  W. Freedman et al., 2001, ApJ, 553, 47-72.

\bibitem{34}  D. Eisenstein et al., 2005, ApJ, 633, 560.

\end{thebibliography}
\end{document}